\documentclass{cpbtex}

\usepackage{graphicx}
\usepackage{dcolumn}
\usepackage{bm}
\usepackage{amssymb}
\usepackage{amsmath}
\usepackage{mathrsfs}

\begin{document}
\begin{CJK*}{GBK}{song}
		
\title{Atomic transport dynamics in crossed optical dipole trap \thanks{Project supported by the National Natural Science Foundation of China (Grants No.~92365208, No.~11934002 and No.~11920101004), the National Key Research and Development Program of China (Grants No.~2021YFA0718300 and No.~2021YFA1400900), the Science and Technology Major Project of Shanxi (Grant No.~202101030201022), and the Space Application System of China Manned Space Program.}}

\author{Peng Peng $^{1}$, \ Zhengxi Zhang $^{1}$, \ Yaoyuan Fan $^{1}$, \\ \ Guoling Yin $^{3}$, \ Dekai Mao $^{1}$, \ Xuzong Chen $^{1}$, \ Wei Xiong $^{1,*}$ \, and Xiaoji Zhou$^{1,2,3,*}$ 
%\author{Peng Peng (彭鹏) $^{1}$, \ Zhengxi Zhang (张正熙) $^{1}$, \ Yaoyuan Fan (樊耀塬) $^{1}$, \\ \ Guoling Yin (殷国玲) $^{3}$, \ Dekai Mao (毛德凯) $^{1}$, \ Xuzong Chen (陈徐宗) $^{1}$, \\ \ Wei Xiong (熊炜) $^{1,*}$ \, and Xiaoji Zhou (周小计) $^{1,2,3,*}$
\thanks{Corresponding author. E-mail:xjzhou@pku.edu.cn}\\
$^{1}${State Key Laboratory of Advanced Optical Communication System and Network,} \\ {School of Electronics, Peking University, Beijing 100871, China}\\
$^{2}${Institute of Carbon-based Thin Film Electronics,} \\ {Peking University, Shanxi, Taiyuan 030012, China}\\
$^{3}${State Key Laboratory of Quantum Optics and Quantum Optics Devices,} \\ {Institute of Opto-Electronics, Shanxi University, Taiyuan 030006, China
}}

\maketitle
\begin{abstract}
We study the dynamical evolution of cold atoms in crossed optical dipole trap theoretically and experimentally. The atomic transport process is accompanied by two competitive kinds of physical mechanics, atomic loading and atomic loss. The loading process normally is negligible in the evaporative cooling experiment on the ground, while it is significant in the preparation of ultra-cold atoms in the space station. Normally, the atomic loading process is much weaker than the atomic loss process, and the atomic number in the center region of the trap decreases monotonically, as reported in previous research. However, when the atomic loading process is comparable to the atomic loss process, the atomic number in the center region of the trap will initially increase to a maximum value and then slowly decrease, and we have observed the phenomenon first. The increase of atomic number in the center region of the trap shows the presence of the loading process, and this will be significant especially under microgravity conditions. We build a theoretical model to analyze the competitive relationship, which coincides with the experimental results well. Furthermore, we have also given the predicted evolutionary behaviors under different conditions. This research provides a solid foundation for further understanding of the atomic transport process in traps. The analysis of loading process is of significant importance for the preparation of ultra-cold atoms in a crossed optical dipole trap under microgravity conditions.
\end{abstract}



\section{Introduction}
The transport of atoms has been a subject of immense interest in recent years\cite{Sarma2011, Argonov2007}, with extensive research being conducted in various fields, such as electronic materials\cite{Roth2009, Konig2007}, trapping ions\cite{Sturm2011, Kielpinski2002}, and cold atoms\cite{Yu2023, Bakhtiari2006} et al. The advantages of using cold atom systems, such as easy manipulation and rapid preparation, make them an attractive option for studying the transport process. Currently, the mainstream atomic transport methods are primarily divided into two categories: magnetic field\cite{Lewandowski2002, Greiner2001, Lewandowski2003} and optical traps\cite{Lee2020, Gross2016, Schmid2006}. Magnetic transport typically relies on the translation of the trap minimum either by mechanically moving a single pair of coils \cite{Lewandowski2002, Pertot2009} or dynamically controlling the current in coils. Nevertheless, optical transport methods are more widely used. The most common implementation is based on a mechanically movable lens that generates a tightly focused optical dipole trap (ODT) with a variable focus position \cite{Lee2020} or a high-precision electrically adjustable optical mount that changes the direction of laser propagation. The transport process of atoms in optical trap has been well studied, such as the transport process from a magneto-optical trap (MOT) to an ODT \cite{Kuppens2000} and the evaporative cooling in ODT and crossed optical dipole traps (CODT)\cite{Davis1995}. 

The CODT is one of the most popular traps in cold-atom experiments. Unlike magnetic traps, the CODT is not selective about the magnetic energy levels of atoms, making it ideal for quantum simulations of different spin states or spinors\cite{Huh2020, Borgh2017, Kawaguchi2012, Tang2019} and quantum precision measurement\cite{Guo2022, Horikoshi2006, Debs2011, Muntinga2013, Dong2021}. The CODT consists of two separate optical traps intersecting at their waists, creating a harmonic oscillator trap in the center and one pair of tubular crossed arms. The transport process of atoms in CODT is accompanied by atomic loss from the trap and atomic loading from the arms to the center region of the CODT. As usual, atoms in the arms will lose quickly, because gravity will tilt the trap and shallow it. The arms of CODT will no longer be able to catch any atoms for their shallower potential depth. The loss process dominates and makes atomic number decrease continually. It is the experimental results observed in previous research, where it just took the loss process into consideration in theoretical models. However, the situation changes under microgravity conditions. When gravity is negligible such as in the Chinese Space Station\cite{Li2023, Li2023a}, the atoms in the arms of CODT will not lose quickly even at the end of evaporative cooling (when the depth of the trap center and the arms are both shallow). The transport of atoms between arms and center has a non-negligible effect on the change of atomic density in the center of the trap and the state of the condensate in the trap. The dynamic properties of evaporative cooling and condensation in space are both different from the experiment on the ground. The research for the transport process in CODT, especially the loading process from arms to center of CODT has a significant value for the preparation of ultra-cold atoms under microgravity conditions and the next research. And that is what we aim to study here.

In this paper, we have conducted a systematic study of the transport process, including the loss process and loading process, in the CODT system. Specifically, we have constructed a set of theoretical models to describe the dynamic evolution of atoms in a CODT. The theoretical models have revealed that the atomic number in the center region of the CODT will decrease monotonically when the atomic loading process is much weaker than the atomic loss process as in previous research. However, when the atomic loading process from the arms to the center is comparable to the atomic loss process, the atomic number in the center region of the CODT will first increase to a maximum value and continually decrease slowly. To validate our theoretical models, we conducted an experiment where we counted the atomic number in the center region of the CODT and the atomic number in the whole trap, respectively. Different from the evaporative cooling, we fixed the trap depth to simulate the situation under microgravity conditions. With fixed large trap depth, the depth in arms is still large enough to hold the atoms for some time despite the presence of gravity. We have observed the increase of atomic number in the center of the trap successfully under the conditions indicated by our theoretical model. This has demonstrated the atomic loading process from arms to the center region in the trap. We then demonstrate that the loading and loss terms are related to the depth of the CODT. The experimental results are consistent with the theoretical model. Moreover, we discuss the evolutionary behaviors of the atomic number in the center region of the CODT under different conditions. It is worth noting that our theoretical model is not only applicable to CODT but also can be applied to other configuration traps. Because of the consideration about the loading process from the arms to the center region, it can be also used to analyze the experiment phenomenon under microgravity conditions. It is highly beneficial to understand the atomic transport process in traps more deeply both on the ground and in the space station.

The remainder of this manuscript is organized as follows. In \textbf{Sec. 2}, we construct a theoretical model of the atomic transport process in a CODT and describe the competitive relationship between the loading and loss processes. The influence of gravity is also discussed. In \textbf{Sec. 3}, we introduce our CODT system and the experimental basis of the transport process. The experimental description and results are given in \textbf{Sec. 4}. Then we extend the evolutionary behavior of atoms in the trap under different conditions in \textbf{Sec. 5} before the conclusions are given in \textbf{Sec. 6}.

\section{Theoretical model}
For a finite-depth trap loaded with atoms, the hotter atoms with high energy will continually escape from the trap, while the cooler atoms with low energy will remain. The loss of hotter atoms is accompanied by the change in the total energy of the remaining trapped atoms, which results in a decrease in the atomic number $N$ and atomic temperature $T$. Meanwhile, the phase space distribution (momentum and density) of cooler atoms also changes and contributes to transport within the trap. Generally, the loss term of atoms is much larger than the loading term, so the density of atoms in the trap decreases everywhere, and the loading term is obscured. However, if the atom density is low and the temperature change is significant, the atom density at the lowest potential position in the trap may increase at certain moments. To analyze the evolution of the atom phase space distribution in a trap in detail, we will discuss a quasi-static approximation result in \textbf{Sec. 2.1} as a qualitative analysis and give a verifiable model for the loss and loading terms in a crossed optical dipole trap in \textbf{Sec. 2.2}. The differences of transport processes in gravity and microgravity environments are also discussed.

\subsection{Quasi-static approximation for evolution of atom distribution}
In the transport process that we focus on, the characteristic time for the change in atomic number is several seconds, while the thermal relaxing time is just several hundred milliseconds. The state of atoms trapped is not a steady state, but it can be considered as a quasi-equilibrium state when we focus on the change of atomic number. Therefore, we take the Boltzmann distribution as an approximation to analyze the evolution of atom distribution qualitatively as shown in Eq.~\ref{eq1}.
\begin{equation}
\label{eq1}
n(\vec{r})=N \frac{{\rm e}^{-\frac{U(\vec{r})}{k_{\rm B} T}}}{ \iiint_V {\rm e}^{-\frac{U(\vec{r})}{k_{\rm B} T}} d^3\vec{r}},  
\end{equation}

\noindent where $k_{\rm B}$ is the Boltzmann constant, and the atomic density $n(\vec{r})$ is determined by the total atomic number $N$, the atomic temperature $T$, and the potential $U(\vec{r})$. By taking the derivative of Eq.\ref{eq1}, we obtain the following Eq.\ref{eq3} as (unit atom number $N'=1$, unit volume $V' = 1$ $\rm{m^3}$)

\begin{equation}
\label{eq3}
{\rm d} n = n  {\rm d} \left[ \ln\left( \frac{N}{N'} \right) \right] + n {\rm d} \left[ \ln \left( \frac{{\rm e}^{-\frac{U(\vec{r})}{k_{\rm B} T}} V'}{\iiint_V {\rm e}^{-\frac{U(\vec{r})}{k_{\rm B} T}} {\rm d}^3\vec{r}}  \right) \right].
\end{equation}

As discussed earlier, the atomic number $N$ and temperature $T$ will continually decrease within limits even for a stationary trap, i.e., $\dfrac{{\rm d} N}{{\rm d} t} <0$ and $\dfrac{{\rm d} T}{{\rm d} t}<0$. Therefore, the first term on the right of  Eq.\ref{eq3} is negative, corresponding to the loss process. Noting that $U(\vec{r})<0$ for a trap, the second term on the right of Eq.\ref{eq3} is positive, which corresponds to the loading process accompanied by the change of $T$.

\subsection{Evolution of atom distribution in crossed optical dipole trap}
\begin{figure*}[htbp]
\includegraphics[width=1\textwidth]{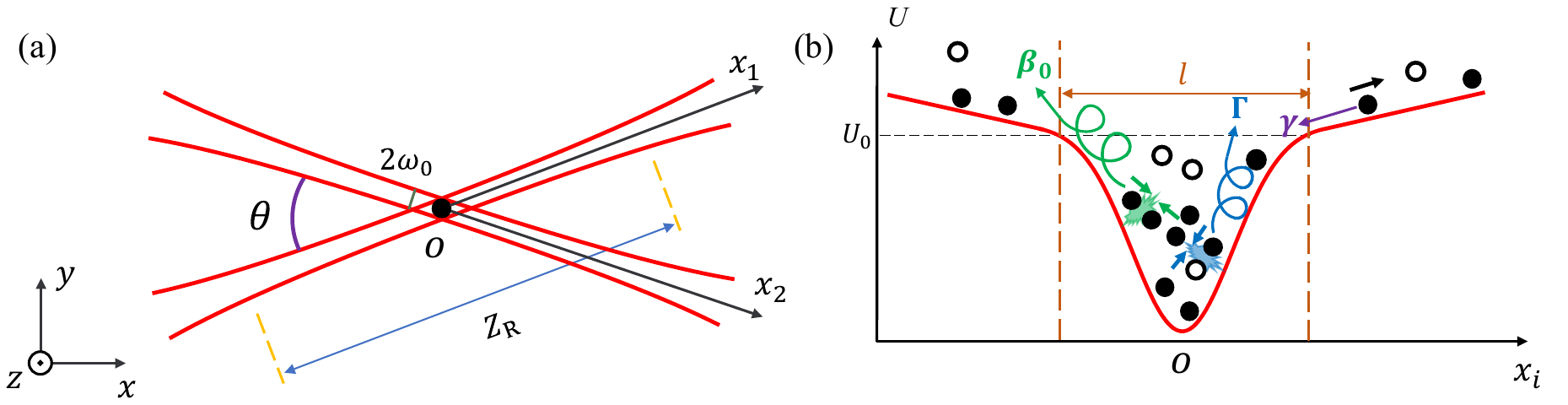}
\caption{(a) The structural schematic diagram of the CODT. $\omega_0$ is the waist radius. $\theta$ is the angle between two beams. $\vec{x}_1$ and $\vec{x}_2$ are the directions of the two ODTs respectively. $\rm Z_R$ is the Rayleigh length of the optical trap. (b) Evolutionary behavior diagram of atoms in a CODT. The solid circles and the hollow circles represent the trapped atoms and background atoms respectively. The red line illustrates the ODT profile along the $\vec{x}_i$ axis, which is the axial direction of the ODT. The trap depth of the CODT in the absence of gravity is $U_{0}$. Due to the interaction between atoms in the CODT and between atoms and the background gas, atoms continuously escape from the trap (collisional loss coefficient $\beta_0$ and exponential loss coefficient $\Gamma$). Simultaneously, atoms trapped in the non-cross part of the optical trap are transported to the center of the CODT (damping coefficient $\gamma$).}
\label{fig1}
\end{figure*}

To analyze the evolution of atom distribution further, we chose the crossed optical dipole trap as an example, which is the same as the experimental setup on the Chinese Space Station. We assume that the CODT consists of two identical Gaussian beams, as shown in Fig.~\ref{fig1}(a). The waist radius of each beam is $\omega_0$ and $Z_R=\dfrac{\pi\omega_0^2}{\lambda}$ represents the Rayleigh length of a single optical trap. $\lambda$ is the wavelength of the optical trap beam. $\theta$ is the angle between two beams. Then the potential can be described by Eq.\ref{eq4} as
\begin{equation}
\label{eq4}
U(\vec{r}) = U(x,y,z) = -\frac{1}{2}k_{\rm B}U_{0} [ \frac{{\rm e}^{-\frac{2(y_1^2+z^2)}{\omega_0^2(1+x_1^2/Z_R^2)}}}{1+x_1^2/Z_R^2}+\frac{{\rm e}^{-\frac{2(y_2^2+z^2)}{\omega_0^2(1+x_2^2/Z_R^2)}}}{1+x_2^2/Z_R^2}] + mgz,
\end{equation}

\begin{equation}
\left\{\begin{array}{c}x_1=\cos(\frac{\theta}{2})x+\sin(\frac{\theta}{2})y, \quad y_1=-\sin(\frac{\theta}{2})x+\cos(\frac{\theta}{2})y 
~\\
\\x_2=\cos(\frac{\theta}{2})x-\sin(\frac{\theta}{2})y, \quad y_2=-\sin(\frac{\theta}{2})x-\cos(\frac{\theta}{2})y \end{array}\right. \nonumber
\end{equation}

\noindent where $U_{0}$ is trap depth of CODT in the absence of gravity, $\vec{x}_1$ and $\vec{x}_2$ are the directions of the two ODTs respectively, $\vec{y}_1$ ($\vec{y}_2$) is orthogonal to $\vec{x}_1$ ($\vec{x}_2$). Due to the presence of gravity, the CODT will be inclined. The effective trap depth $U_{\rm eff}$ can be calculated by Eq.~\ref{eq4}. Similarly to the expression in Eq.~\ref{eq3}, we can obtain the change of the atomic number in the center region of the CODT by Eq.~\ref{eq6}

\begin{equation}
\label{eq6}
\frac{{\rm d} N_c}{{\rm d}t}= R(t)-D(t),
\end{equation}

\noindent where $N_c$ is the atomic number in the center region of the CODT, $D(t)$ represents the loss term, and $R(t)$ represents the loading term. Detailed expressions for these two components will be provided below.

\subsubsection{Loss process}
The loss of atoms originates from the escape of hotter atoms. Therefore, the loss rate is determined by the generation rate of hotter atoms, which is primarily induced by collisions between trapped atoms or between trapped atoms and background atoms for a far-detuned ODT. Generally, the loss part related to the density can be approximately given by \cite{Bismut2011, Shibata2017, Schlosser2002, Barker2023, Mies1996, Miller1993, Wang2023}
\begin{equation}
\label{eq7}
\frac{{\rm d} n}{{\rm d}t}=-\Gamma n - \beta_0 n^2.
\end{equation}

\noindent As shown in Fig.~\ref{fig1}(b), $\Gamma$ is the exponential loss rate corresponding to collisions between trapped atoms and background atoms, while $\beta_0$ is the collision loss coefficient which is related to the atomic density between trapped atoms themselves. 

The collisions between trapped atoms and background atoms are affected by their densities and temperatures. However, the temperature of the background atoms $T_b$ is much higher than that of the trapped atoms, i.e. the mean velocity of background atoms $\overline{v}_b$ is much larger than the mean velocity of trapped atoms $\overline{v}$. Both the collision rate and escape possibility of trapped atoms after collisions primarily depend on the background atoms. What's more, $T_b$ is also much higher than trap depth $U_{\rm eff}$ in general. As a result, we can approximate that almost every collision between trapped atoms and background atoms leads to a trapped atom being lost. So we have the exponential loss rate $\Gamma$ as,
\begin{equation}
\label{eqGamma}
\Gamma=\sigma n_b \overline{v}_b,
\end{equation}
\noindent where $\sigma$ is the scattering cross-section of atoms, $n_b$ is the density of background atoms. There are almost no other gas atoms except Rubidium because the experimental system is in an ultra-high vacuum, $(1.3 \pm 0.3)\times10^{-9}$ Pa. The background atoms are high-energy Rubidium atoms that are not trapped. Both background atoms and trapped atoms have mass $m$. The mean velocity of background atoms $\overline{v}_b$ can be taken as $\overline{v}_b=\sqrt{\frac{8 k_{\rm B} T_b }{\pi m}}$.

The collisions between trapped atoms have been extensively studied\cite{Kuppens2000, Davis1995}. This kind of collision loss in the far-detuned trap is due to both photoassociation and ground state hyperfine changing collisions, which does not strongly depend on the trap depth or the wavelength of the optical trap beam. So the loss coefficient of collisions between trapped atoms $\beta_0$ can be expressed as below, 
\begin{equation}
\label{eqbeta}
\beta_0= \sqrt{2} \sigma \overline{v} P,
\end{equation}
\noindent where $\sqrt{2} \sigma \overline{v}$ is the collision rate, and $P$ represents the loss probability for one collision.

By integrating Eq.~\ref{eq7} over a selected space, such as the center region of the CODT, we obtain the loss rate of $N_c$ as follows
\begin{equation}
\label{eq8}
\frac{{\rm d} N_c}{{\rm d} t}=-\Gamma N_c - \frac{\beta_0}{V_c} N_c^2, \quad  V_c=\frac{\left(\iiint n d^3 \vec{r}\right)^2}{\iiint n^2 d^3 \vec{r}},
\end{equation}

\noindent where the effective volume $V_c$ is determined by the spatial distribution of atoms. Thus, the loss term $D(t)$ can be expressed as in Eq.~\ref{eq10}
\begin{equation}
\label{eq10}
D(t)=\Gamma N_c + \beta N_c^2, \quad  \beta=\beta_0/V_c.
\end{equation}
\noindent where $\beta$ is the collision loss coefficient related to $V_c$.

It is worthing that the model stated above is just effective for the trapped atoms. When there is no trap, such as the arms of over-tilted CODT on the ground, the atoms will lose rapidly and can not be described with Eq.~\ref{eq10}.

\subsubsection{Loading process}
\begin{figure*}[htbp]
\centering
\includegraphics[width=1\textwidth]{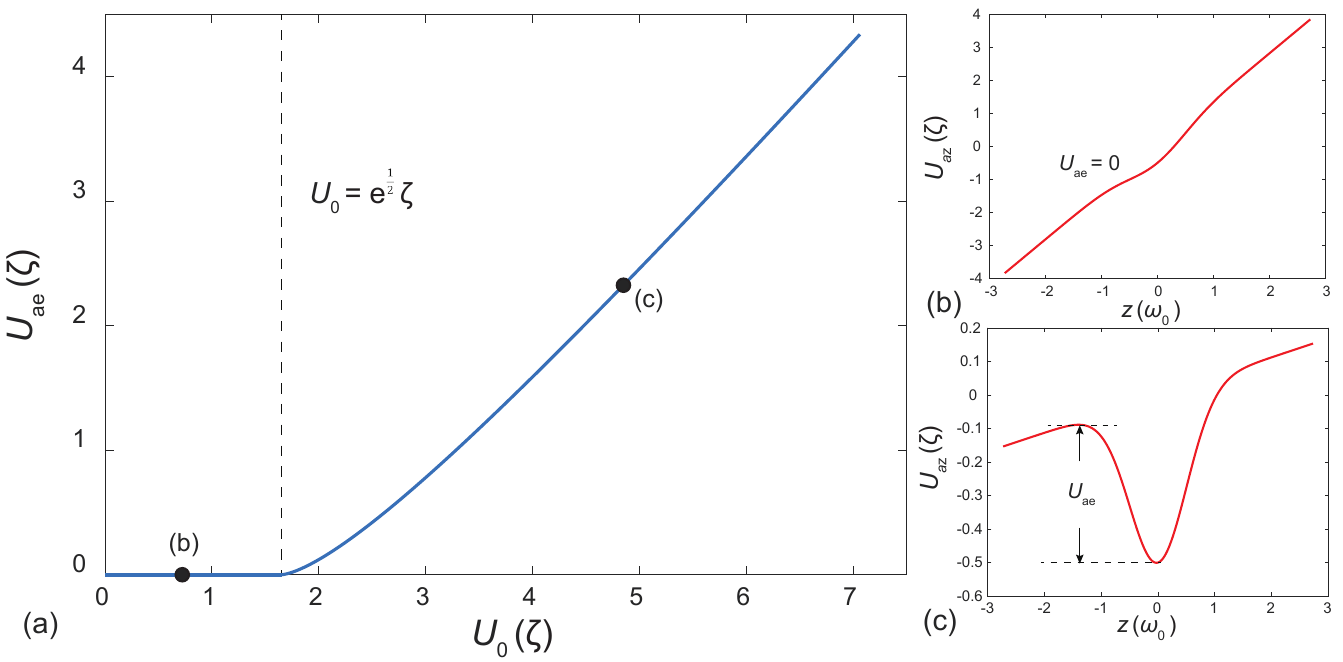}
\caption{(a) The effective potential depth in the arms $U_{\rm{ae}}$ as a function of the trap depth $U_{0}$. $U_{0}$ has a critical value $\rm{e}^{\frac{1}{2}} \zeta$, where $\zeta = mg\omega_0/k_{\rm{B}}$. (b-c) The trap potential energy in the arms along the gravitational direction (along $\vec{z}$) $U_{\rm{az}}$. (b) When $U_0 \leq \rm{e}^{\frac{1}{2}} \zeta$, the arms of CODT can't trap any atom. (c) When $U_0>\rm{e}^{\frac{1}{2}} \zeta$, the arms of CODT can trap atoms despite the trap is tilted.}
\label{figeff}
\end{figure*}

The loading process is essentially the transfer of atoms from a higher potential position to a lower potential position. However, in the experiment on the ground, due to gravity, the potential will become shallow, and the atoms at the higher potential will lose if there is no trap. This results in almost no loading process, which has happened in previous research, especially in evaporative cooling. For a CODT, similar to the calculation of effective trap depth $U_{\rm{eff}}$, we can calculate the effective potential depth in the arms $U_{\rm{ae}}$. The results are shown in Fig.~\ref{figeff}, where $\zeta = mg\omega_0/k_{\rm{B}}$. From the Fig.~\ref{figeff}, we can see there will be no trap when $U_0 \leq \rm{e}^{\frac{1}{2}} \zeta$. In fact, the critical value will be higher than $\rm{e}^{\frac{1}{2}} \zeta$ because atoms can't be cooled to absolute zero. With slight loading process and severe loss process, the atomic number in the center of the CODT will decrease monotonically, and the loading process has been reasonably ignored in previous research. However, under microgravity conditions, with no tilt of the trap, the loading process from the arms of CODT to the center will exist persistently, even if after evaporative cooling (It has been observed in the Chinese Space Station, and will be discussed in the future work). And there is a need to take the loading process into consideration to analyze the transport process in CODT accurately. As a general analysis, we have constructed a model for the loading process.

Taking the loss process and loading process into consideration separately, the loading process of trapped atoms is related to the change in atom energy, which is directly proportional to the atomic temperature $T$. The motion of atoms in the center of the trap resembles oscillation accompanied by collisions. For a collision resulting in atom loss, the energy of atom remaining in the trap will be carried away by atom expelled. This can be treated as a damping force on the atom. By setting the damping coefficient as $\gamma$, the decay rate of the atomic motion will be ${\rm e}^{-\gamma t}$ similar to damped oscillations approximately. So as an estimation of the number at which atoms outside the trap center enter the center area and are trapped per unit of time, the loading term $R(t)$ for the atoms in the center of the trap can be expressed as
\begin{equation}
\label{eq12}
R(t)= 2 \gamma N_0 f(t) {\rm e}^{-f(t)}, f(t)={\rm e}^{2 \gamma t},
\end{equation}

\noindent where the damping coefficient $\gamma$ is determined by the change in temperature $T$, and $N_0$ is the total atomic number at $t=0$. Based on the quasi-static approximation for the evolution of atom distribution, the relationship between the damping coefficient $\gamma$ and temperature change can be expressed as
\begin{equation}
\label{eq15}
\gamma=-\frac{1}{T}\frac{{\rm d} T}{{\rm d} t}.
\end{equation}

\noindent The quasi-static approximation we take here means that the distribution of atoms is considered to be close to the Boltzmann distribution. The energy of one atom in the optical trap is taken as $3 k_{\rm B} T$ approximately (the average energy of a harmonic oscillator).

\subsubsection{Two possible evolutionary behaviors}
When considering both the loss and loading processes, $\dfrac{{\rm d} N_c}{{\rm d}t} $ can be expressed as
\begin{equation}
\label{eq16}
\frac{{\rm d} N_c}{{\rm d} t} = - D(t) + R(t) = -\Gamma N_c -\beta N_c^2 +2\gamma N_{0} f(t) {\rm e}^{- f(t)}.
\end{equation}

\noindent The loading term $R(t)$ is negative with respect to $t$. As a result, $N_c$ will ultimately continue to decrease when time is sufficiently long, even if $N_c$ increases at the beginning. When $\dfrac{{\rm d} N_c}{{\rm d} t}=0$, since $R(t)$ is negative with respect to $t$ and $D(t)$ is positive with respect to $N_c$, the value of $N_c$ can only decrease. Therefore, there are two cases:
\begin{align*}
&\mathrm{I.}&&\left( \frac{{\rm d} N_c}{{\rm d} t} \right)_{t=0} \leq 0 \quad \Rightarrow \quad \frac{{\rm d} N_c}{{\rm d} t} \leq 0;\\
&\mathrm{II.}&&\left( \frac{{\rm d} N_c}{{\rm d} t} \right)_{t=0} >0 \quad \Rightarrow \quad \frac{{\rm d} N_c}{{\rm d} t} (t-t_m) \leq 0, \quad \left( \frac{{\rm d} N_c}{{\rm d} t} \right)_{t=t_m} =0.
\end{align*}

\noindent where $t=t_m$ corresponds to the maximum point of $N_c$ for case $\mathrm{II}$. In case $\mathrm{I}$, the loading term is less than the loss term at all times, and $N_c$ will continuously decline. Remarkably, in case $\mathrm{II}$, $N_c$ will increase non-trivially at first, and the condition is just $\left( \dfrac{{\rm d} N_c}{{\rm d} t} \right)_{t=0} > 0$. At $t=0$, we have
\begin{equation}
\label{eq17}
\left( \frac{d N_c}{d t} \right)_{t=0} = -\Gamma N_{c0} -\beta N_{c0}^2 +2\gamma N_{0} {\rm e}^{-1},
\end{equation}
where $N_{c0}$ and $N_0$ are the atomic numbers in the center of the CODT and the total atomic number at $t=0$, respectively. Noting that $N_{c0}$ is proportional to $N_0$, we define that $\alpha=\dfrac{N_{c0}}{N_0}$. It is evident that $\left( \dfrac{{\rm d} N_c}{{\rm d} t} \right)_{t=0}$ will be negative if $N_0$ is sufficiently large. The condition for case $\mathrm{II}$, $\left( \dfrac{{\rm d} N_c}{{\rm d} t} \right)_{t=0} >0$, requires that
\begin{equation}
\label{eq18}
N_{c0} < \frac{1}{\alpha \beta} (-\Gamma \alpha+2\gamma {\rm e}^{-1}).
\end{equation}

% \noindent and
% \begin{equation}
% \label{eq19}
% \frac{2 \gamma}{\alpha \Gamma e} >1.
% \end{equation}
Hence it is expected that a nontrivial increase of $N_c$ will be observed at low atom density $N_{c0}$, where there will be a better platform to research the transport process.

\section{System setup}

To study the above processes, our experiment begins with a magneto-optical trap (MOT), which cools and traps ${\rm ^{87}Rb}$ atoms from an ultrahigh vacuum of $(1.3 \pm 0.3)\times10^{-9}$ Pa. Two low-power extended cavity diode lasers (ECDL) (DL 100, Toptica, Germany) are used for the cooling and repump beams. We lock the repump beams using the modulation transfer spectroscopy (MTS) technique which is resonant to the $5^2 S_{1/2}$ F = 1 to $5^2 P_{3/2}$ F = 2 transition. The cooling beams are locked using the beat frequency phase-locking (BFPL) technique, red detuned by -18 MHz from the $5^2 S_{1/2}$ F = 2 to $5^2 P_{3/2}$ F' = 3 transition. The waist radius of the beams is 10 mm. The intensities of the cooling and repump beams are $I_{\rm c}=5.2$ $\mathrm{mW/cm^2}$ and $I_{\rm r}=0.4$ $\mathrm{mW/cm^2}$, respectively. The quadrupole field along the strong axis has a magnetic field gradient of 8 $\mathrm{Gauss/cm}$. About $7\times10^8$ atoms are collected in the MOT within 5 seconds.

\begin{figure*}[htbp]
\includegraphics[width=1\textwidth]{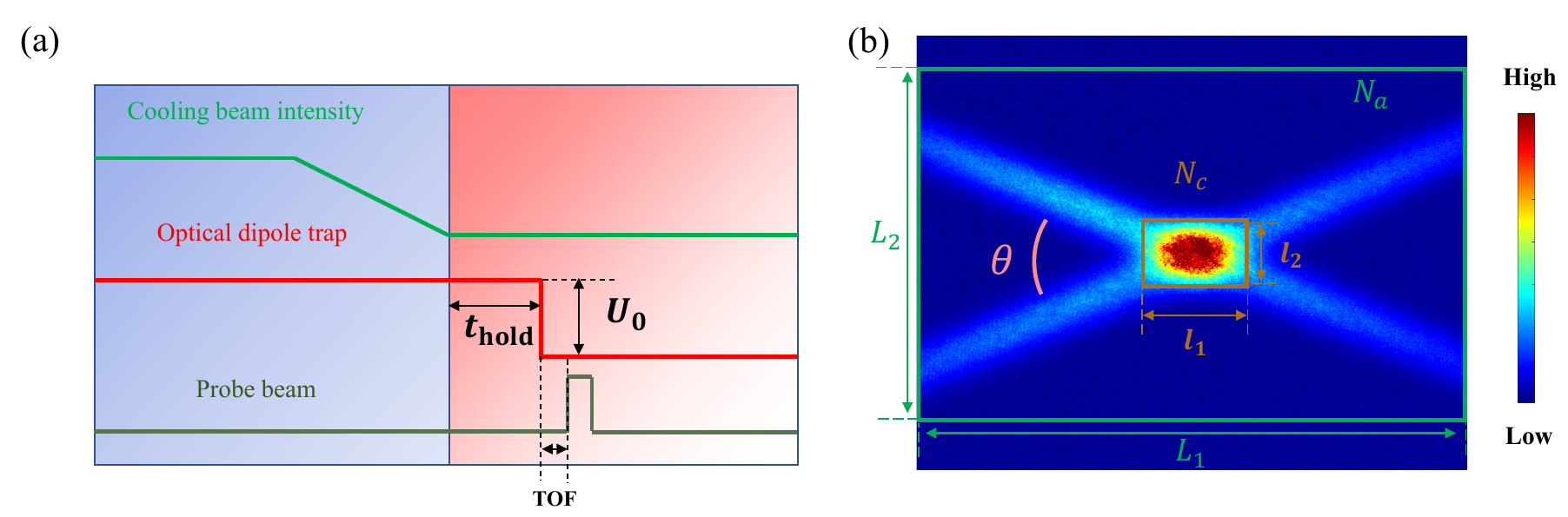}
\caption{(a) The time sequence diagram of the experiment. The holding time of the CODT is $t_{\rm hold}$. (b) The area within the brown rectangle (green rectangle) represents the central area of the trap (size with molasses). The atomic number in the brown rectangle (green rectangle) is $N_c$ ($N_a$). The total atomic number $N_a$ at $t=0$ is just $N_0$ in Eq. \ref{eq12}. The two sides of the brown rectangle are $l_1=6\omega_0$ and $l_2=4.5\omega_0$ respectively, and the atomic density on the side of the rectangle is almost ${\rm e}^{-1}$ of the atomic density of the center of the CODT. The sides of the green rectangle correspond to the area where molasses covered are $L_1$ and $L_2$ respectively. }\label{fig2}
\end{figure*}

After loading a sufficient number of atoms into the MOT, we increase the atomic density by employing the Dark-MOT process. The process reduces the intensity of the repump beams and increases the detuning of the cooling beams to force cold atoms into a dark state. To decrease the temperature of the atoms, the detuning of the cooling beams is swept from -30 MHz to -70 MHz continuously over 9.0 ms. Meanwhile, we decrease the intensity of the cooling beams to 0.8 mW/cm$^2$. Then, we maintain the repump beams for an additional 3.0 ms to ensure that almost all atoms are pumped to the $5^2 S_{1/2}$ F = 1 state. The atomic number and temperature are measured through absorption imaging after the time of flight (TOF). As a result of the process, the temperature of atoms can drop to 40 $\mathrm{\mu K}$.

We capture atoms from optical molasses by CODT. The wavelength of the optical trap beam is 1064 nm, which is far red-detuned from the resonant frequency of ${\rm ^{87}Rb}$ atoms. The waist radius of the optical trap beam $\omega_0$ is 55 $\mathrm{\mu m}$, and the included angle between the two beams $\theta$ is 45 degrees. The two beams are focused on the center of the molasses. To avoid the formation of an optical lattice, the polarization of the two optical traps is orthogonal. Optical trap powers are controlled by a dual detector power feedback system, which reduces the power instability to 0.15 $\%$. The experimental sequence is shown in Fig.~\ref{fig2}(a). 

The CODT is turned on at the start of the MOT. The cooling and repumping beams are turned off after the molasses process. After the holding atoms $t_{\rm hold}$  in the CODT, the atoms that are out of the trap will be separated from the atoms in the trap. As shown in Fig.~\ref{fig2}(b), the range of $N_a$ corresponds to the area where molasses is covered, and the range of $N_c$ is chosen as a rectangle with $l_1$ and $l_2$ which has the center the same center as the CODT. The sides $l_1$ and $l_2$ are selected as $6\omega_0$ and $4.5\omega_0$ so that the atomic density on the sides of the rectangle is ${\rm e}^{-1}$ of the atomic density of the center of the CODT.

\section{Experimental results}
As discussed above, to directly observe the atomic transport phenomena in the trap, we load atoms into the CODT with different initial mean densities by adjusting the experimental parameters of the Dark-MOT and the molasses processes. The initial central atomic number $N_{c0}$ and initial total atomic number $N_0$ are both proportional to the initial mean density, and they directly affect the evolution of the atomic distribution as analyzed in section Sec. 2.2.3. As a result, different evolution phenomena are expected to occur and have been observed. 

\begin{figure*}[htbp]
	\includegraphics[width=1\linewidth]{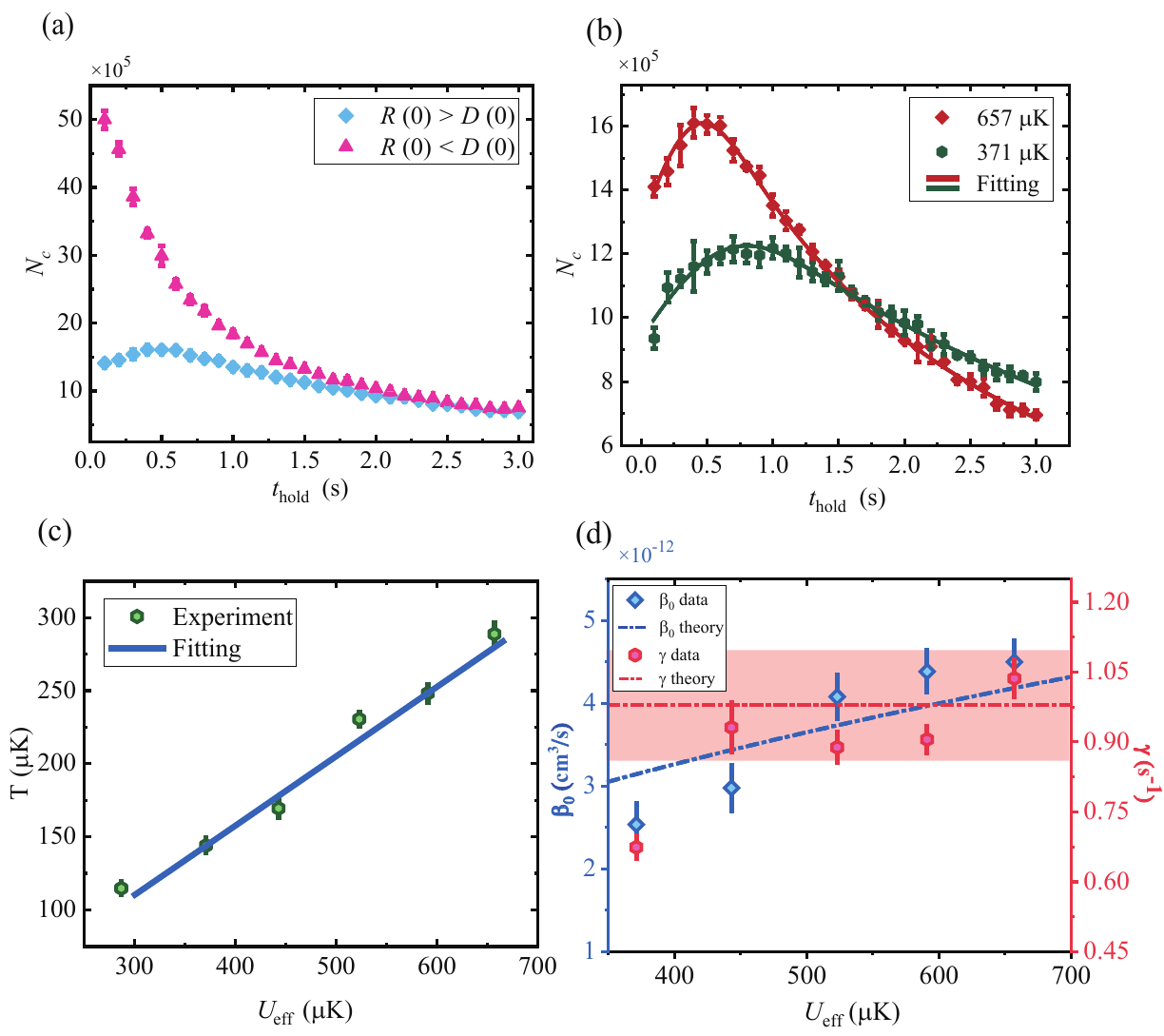}
	\caption{(a) $N_c$ in the CODT as a function of $t_{\rm hold}$.  Pink triangle data represent that the atomic loading effect is much weaker than atomic loss process ($R(0) < D(0)$). And $N_c$ in the CODT decreases monotonically. Blue rhombus data represent that the atomic loading process is stronger than the atomic loss process at the beginning. ($R(0) > D(0)$). $N_c$ first increases to a maximum value and then slowly decreases. The trap depths of two cases are 1249 $\mu$K and 657 $\mu$K, and the initial atomic number $N_{c0}$ of two cases are $5.02 \times 10^6$ and $1.41 \times 10^6$. (b) The relationship between the center atomic number $N_c$ in the CODT and $t_{\rm hold}$. The solid lines represent the results of the theoretical fitting according to Eq.\ref{eq16}. The fitting parameters include $\gamma$ and $\beta_0$. The $\Gamma$ is calculated with Eq. \ref{eqGamma}. The initial atomic number $N_0$ and $N_{c0}$ are measured experimentally.  Each data point is the average result of five measurements. The orange diamonds (green hexagons) represent the CODT with a trap depth of 657 $\mu$K (371 $\mu$K). (c) Temperature $T$ of the atoms in the CODT as a function of trap depth $U_{\rm eff}$, for $\omega_0$ = 55 ${\rm \mu m}$. The solid line is a linear fitting, the slope $\eta = T/U_{\rm eff}$ = 0.475 $\pm$ 0.032. (d) Collisional loss coefficient $\beta_0$ and damping coefficient $\gamma$ over trap depths. The blue rhombuses (red hexagons) are experimental data of collisional loss coefficient $\beta_0$ (damping coefficient $\gamma$). The error bars of the data represent the standard deviation of the five measurements. The dashed line is the result of theoretical calculation according to Eq.\ref{eq16}. The light red area represents the estimation interval of $\gamma$.}\label{fig3.1}
\end{figure*}

Fig.~\ref{fig3.1}(a) shows the results of the experiment: the evolution of $N_c$ during $t_{\rm hold}$ for case $\mathrm{I}$ which is represented by red triangles and case $\mathrm{II}$ which is represented by blue rhombus. In case $\mathrm{I}$, with large initial atom number $N_{c0}$, the loss term $D(t)$ of the physical system is larger than the loading term $R(t)$ at the beginning, $R(0) < D(0)$, and the atomic number $N_c$ decreases continuously all the time, $\left(\dfrac{{\rm d} N_c}{{\rm d} t}\right)_{t=0} <0$. Meanwhile in case $\mathrm{II}$, $R(0) > D(0)$ and $\left(\dfrac{{\rm d} N_c}{{\rm d} t}\right)_{t=0} >0$. When the initial atomic density is low enough, as shown by the blue rhombus, the central atomic number $N_c$ increases rapidly to a maximum first and then decreases similarly to the situation in case $\mathrm{I}$. The nontrivial increase of $N_c$ demonstrates the presence of transport from the tubes of CODT directly to the center. The experimental results observed are consistent with the theoretical analysis in Sec. 2.2. 

To understand the relation between the loss process and loading process deeper, we further measured the evolution of $N_c$ during the holding time $t_{\rm hold}$ in the CODT with different trap depths $U_{\rm eff}$ in case $\mathrm{II}$. Representative experimental results in deep (657 $\mu$K) and shallow (371 $\mu$K) traps are shown in Fig.\ref{fig3.1}(b). Similar evolution phenomena have been observed, especially the nontrivial increase at the beginning. The experimental conditions before the holding time are all the same except for the trap depth. A deeper trap can load more atoms, and the change rate of $N_c$ is also larger due to the larger $N_{c}$(see Eq.\ref{eq16}). As shown in Fig.\ref{fig3.1}(b), $N_c$ reaches a maximum value of $1.61 \times 10^6$ after about 0.5 s when the trap depth of the CODT is 657 $\mu$K. Meanwhile, the time required to reach the maximum point, when the trap depth of the CODT is 371 $\mu$K, is about 0.9 s. And the maximum value of $N_c$ can only reach about $1.21 \times 10^6$. In short, with deeper trap depth, the peak point of $N_c$ is higher and comes earlier. Significantly, the atomic number $N_c$ in the deeper trap with larger $N_{c0}$ is smaller in the final, which indicates the collisional loss coefficient related to density $\beta_0$ is larger in deeper trap . 

According to Eq.\ref{eq16}, the evolution of $N_c$ is entirely determined by $\Gamma$, $\beta_0$, $\gamma$, $N_0$, and $N_{c0}$. The first one $\Gamma$ is calculated by Eq.\ref{eqGamma}. According to the temperature and density of background atoms measured, we have $\Gamma= (0.068 \pm 0.016 )$ $\rm {s^{-1}}$, which agrees well with the value range in existing research\cite{Kuppens2000}. The uncertainty of $\Gamma$ is calculated from the uncertainty of $T_{b}$ and $n_{b}$. The parameter $\beta_0$ and $\gamma$ are both related to temperature, and the change in atomic density is mainly determined by the change in atomic number and temperature(Sec. 2). So we measured the temperature of atoms trapped at different trap depths. We found the ratios of temperature $T$ to trap depth $U_{\rm eff}$ for different trap depths are almost the same. The ratios at $t=0$ is shown in  Fig.\ref{fig3.1}(c), where we define

\begin{equation}
\label{eqata}
\eta=T/U_{\rm eff}.
\end{equation}

\noindent The relationship between $T$ and $U_{\rm eff}$ is near to be proportional, and the proportionality factor $\eta = $ 0.475 $\pm$ 0.032 over the range of trap depths investigated.

As shown in Fig.\ref{fig3.1}(d), we obtained the collisional loss coefficient related to density $\beta_0$ (blue rhombuses) and energy dissipation factor $\gamma$ (red hexagons) at five different trap depths by fitting the experimental data according to Eq.\ref{eq16}. The energy dissipation factor $\gamma$ is related to the change rate of temperature. Since the proportionality factor $\eta$ is almost the same for different trap depths $U_{\rm eff}$ at different $t_{\rm {hold}}$, $\gamma$ is expected to be a constant approximately. The red dashed line is the result of theoretical calculation for $\gamma$ using temperature measured based on Eq.\ref{eq16}, which corresponds to $\gamma = 0.979$ $\rm{s^{-1}}$. The light red area represents the estimation interval of $\gamma$ by temperature measurement, which covers the experimental fluctuation range for different trap depths. When $U_{\rm eff}=371$ ${\rm \mu K}$, the value of $\gamma$ is less than the theoretical calculation result. That is because $\eta$ is not strictly the same for different $U_{\rm eff}$ all the time. As the trap depth decreases, the change rate of $\eta$ over time also slightly decreases (the range of variation is comparable to the uncertainty of the proportionality factor $\eta$), and makes the value of $\gamma$ in shallow trap smaller. The values of collisional loss coefficient related to density $\beta_0$ in a far-detuned ODT have been calculated through the equation $\beta_0=\beta V_c$. The $\beta$ is obtained by fitting, and the effective volume $V_c$ is estimated by integrating for the central area of the trap in Fig.\ref{fig2}(b) based on quasi-static approximation. And the results of $\beta_0$ (blue rhombuses) are in agreement with the findings of \cite{Kuppens2000} which reported the value of $(4 \pm 2) \times 10^{-12} \rm {cm^3 s^{-1}}$. The blue dash line is the theoretical curve for $\beta_0$ based on Eq.\ref{eqbeta}. The monotonic increase of $\beta_0$ with the trap depth is consistent with the physics regime analyzed. And the larger value of $\beta_0$ for the deeper trap agrees with the discussion about the phenomenon mentioned above (The atomic number is relatively smaller in the deeper trap at the end in Fig.\ref{fig3.1}(b)). 

The deviation of experimental results from theory may be due in part to errors in measuring atomic number and temperature during the experiment. Meanwhile, the quasi-static approximation does not accurately describe the evolution of the atom distribution. The evolution process is a non-equilibrium process strictly. Especially when the trap depth is shallow, the CODT cannot hold atoms tightly, which makes the value of $\beta_0$ lower than the theoretical curve in Fig.\ref{fig3.1}(d). Because of the gravity, the effective potential depth in the arms $U_{\rm{ae}}$ is smaller, and it will decrease the loading process of atoms from the arms of CODT to the center. This is also one possible reason to cause the deviation of $\gamma$ from the theoretical value in the shallow trap.

\section{Discussion}

\begin{figure}[htbp]
\centering
\includegraphics[width=0.6\linewidth]{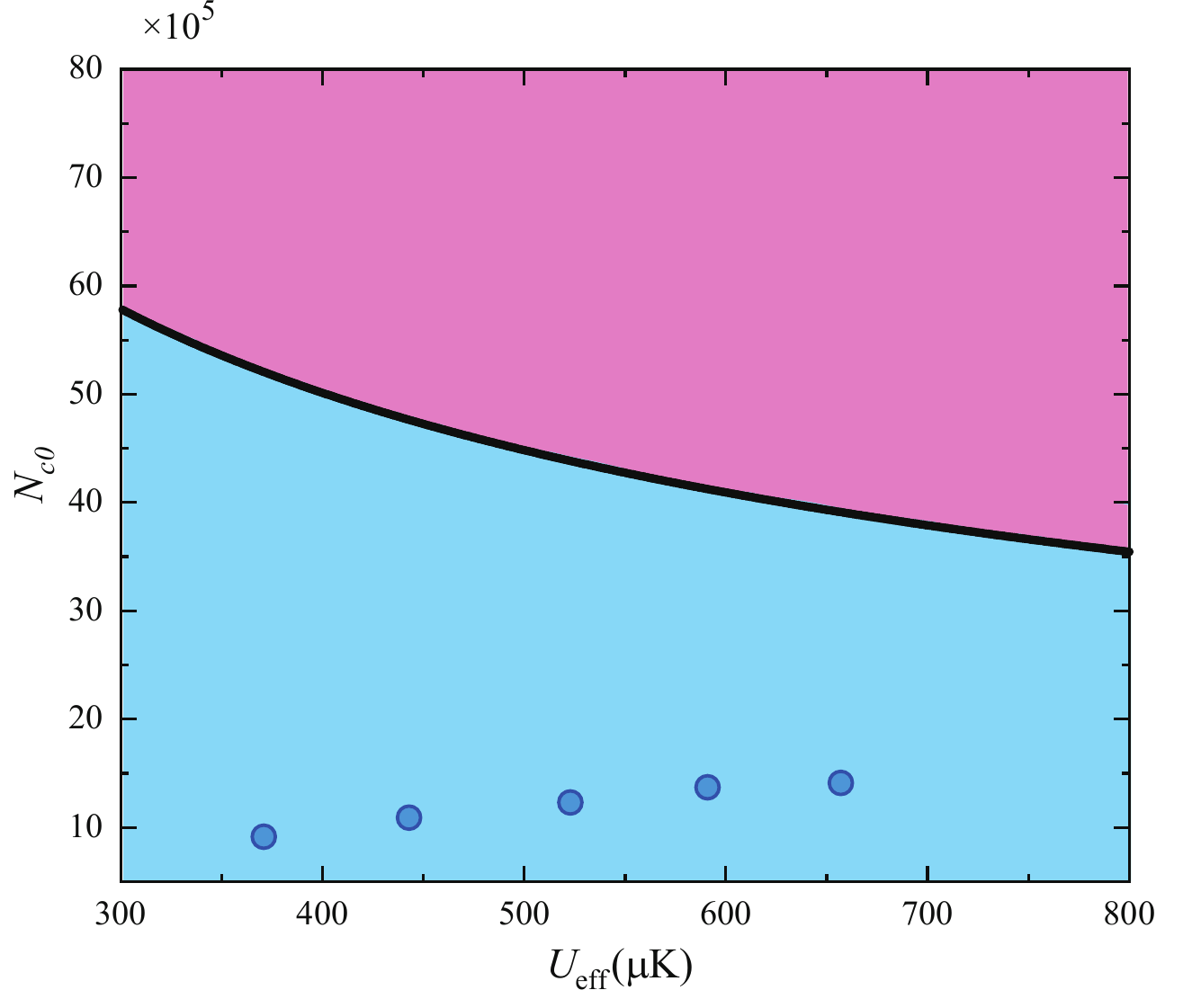}
\caption{The predicted evolutionary behaviors of atomic distribution in CODT under different experimental conditions (different initial central atomic number $N_{c0}$ and effective trap depth $U_{\rm{eff}}$). The pink area represents case I:$R(0) < D(0)$, and the loss term of atoms in the trap will dominate, leading to a continuous decline in $N_c$. The light blue area represents case II: $R(0) > D(0)$, and $N_c$ rapidly will increase to a maximum value initially and then decrease. The five blue circles are the experimental data points in Fig.\ref{fig3.1}(d). The black curve is the dividing line between the two evolutionary behaviors,where $R(0) = D(0)$.}\label{fig6}
\end{figure}

According to Eq.\ref{eq18}, we can obtain the evolutionary behavior of atom distribution in the CODT under more varied conditions. For the same experiment parameters in Sec. 3, we assume $\alpha(\eta)$, $\Gamma$ and $\gamma$ to be fixed values like the experimental results shown above ($\Gamma = 0.068$ $\rm {s^{-1}}$,$\gamma = 0.979$ $\rm {s^{-1}}$). The $\alpha$ is the ratio of the atomic number in the center of CODT to the total atomic number at the beginning. In the quasi-static approximation, a fixed $\eta$ leads to fixed $\alpha$, and $\alpha$ is measured as 0.658. We consider two experimental conditions: trap depth $U_{\rm {eff}}$ and the initial central atomic number $N_{c0}$ in CODT, as independent variables to predict the changes in the loading term $R(t)$ and the loss term $D(t)$ of the atomic system. According to the discussion in Sec. 2.2.3, we have calculated the size relationship between $R(0)$ and $D(0)$, and obtained the predicted evolutionary behaviors of atomic distribution based on the theoretical model with different $N_{c0}$ and $U_{\rm {eff}}$. The result is shown in Fig.\ref{fig6}. In the pink area, where $R(0) < D(0)$ is satisfied, the evolutionary behavior belongs to case I: the loss process of atoms in the trap dominates, and leads to a continuous decline in $N_c$. Conversely, in the light blue area, where $R(0) > D(0)$ holds, the evolutionary behavior belongs to case II: $N_c$ rapidly increases to a maximum value at first and then decreases. The five blue circles represent the experimental data for five different trap depths shown in Fig.\ref{fig3.1}(d). All of them fall within the light blue region, indicating that the evolutionary behaviors should belong to case II. The experimental results of the atomic number evolution have indeed shown an initial increase, which agrees well with the predictions. The consistency between the predictions based on our theoretical model and experimental results has shown the validity of our model for understanding the transport process in the CODT.
% The point for red triangle data shown in Fig.\ref{fig3.1}(a) has not been drawn in the Fig.\ref{fig6}. It is in the pink area, which is also consistent with the experiment result.
%When the change in $\gamma$ and $\eta$ is small, larger $N_c$ leads to smaller $t_m$ for $ \left( \frac{d N_c}{d t} \right)_{t=t_m}=0$ according to Eq.\ref{eq16}. That's why the maximum point of $N_c$ for the deeper trap with larger $N_c$ corresponds to the shorter holding time in Fig.\ref{fig3.1}(b).

\section{Summary}
In summary, we have systematically studied the atomic transport process in a crossed optical dipole trap (CODT). Atoms in the CODT undergo both loading and loss processes, between which there is a competitive relationship. For the shallow trap on the ground, the loading process between different parts of the trap can be ignored. However, under microgravity conditions, the loading process persists and must be taken into account. When the atomic loss term in the trap dominates, $N_c$ declines monotonically. Conversely, when the atomic loading process is comparable to the atomic loss process, $N_c$ rapidly increases to a maximum value initially and then decreases. Our experimental results verify the competition mechanism between the two physical processes and explore the relationship between these processes and the trap depth. This highlights the complexity of atomic dynamic evolution in the trap. The experimental and theoretical results are in good agreement with each other. Our research advances the understanding of the atomic transport process in CODT, which is significant under microgravity conditions like the Chinese Space Station.

% \bibliography{Ref.bib}

\end{CJK*} 
\end{document}